\documentclass[%
 aip,
 amsmath,amssymb,
 reprint,%
]{revtex4-1}

\usepackage{graphicx}
\usepackage{dcolumn}
\usepackage{bm}

\usepackage[utf8]{inputenc}
\usepackage[T1]{fontenc}
\usepackage{mathptmx}
\usepackage{etoolbox}
\usepackage[per-mode=repeated-symbol]{siunitx}
\usepackage{array}
\usepackage[mathscr]{euscript}

\makeatletter
\def\@email#1#2{%
 \endgroup
 \patchcmd{\titleblock@produce}
  {\frontmatter@RRAPformat}
  {\frontmatter@RRAPformat{\produce@RRAP{*#1\href{mailto:#2}{#2}}}\frontmatter@RRAPformat}
  {}{}
}%
\makeatother
\begin{document}

\preprint{AIP/123-QED}

\title[Energy scaling in a compact bulk multi-pass cell enabled by Laguerre-Gaussian single-vortex beams]{Energy scaling in a compact bulk multi-pass cell \\ enabled by Laguerre-Gaussian single-vortex beams}
\author{V. Koltalo\textsuperscript{*}}
 \email[]{victor.koltalo@ensta-paris.fr, anne-lise.viotti@fysik.lth.se}
 \affiliation{Laboratoire d’Optique Appliquée (LOA), Institut Polytechnique de Paris, ENSTA Paris - CNRS - Ecole Polytechnique, 91120 Palaiseau, France}

\author{S. Westerberg}
 \affiliation{Department of Physics, Lund University, P.O. Box 118, SE-22100 Lund, Sweden}
\author{M. Redon}
 \affiliation{Department of Physics, Lund University, P.O. Box 118, SE-22100 Lund, Sweden}
\author{G. Beaufort}
 \affiliation{Department of Physics, Lund University, P.O. Box 118, SE-22100 Lund, Sweden}
\author{A.-K. Raab}
 \affiliation{Department of Physics, Lund University, P.O. Box 118, SE-22100 Lund, Sweden}
\author{C. Guo}
 \affiliation{Department of Physics, Lund University, P.O. Box 118, SE-22100 Lund, Sweden}
\author{C. L. Arnold}
 \affiliation{Department of Physics, Lund University, P.O. Box 118, SE-22100 Lund, Sweden}
\author{A.-L. Viotti\textsuperscript{*}}
 \affiliation{Department of Physics, Lund University, P.O. Box 118, SE-22100 Lund, Sweden}
\date{\today}

\begin{abstract}

We report pulse energy scaling enabled by the use of Laguerre-Gaussian single-vortex ($\text{LG}_{0,l}$) beams for spectral broadening in a sub-40 cm long Herriott-type bulk multi-pass cell. Beams with orders ${l= 1-3}$ are generated by a spatial light modulator, which facilitates rapid and precise reconfiguration of the experimental conditions. $\SI{180}{\femto\second}$ pulses with $\SI{610}{\micro\joule}$ pulse energy are post-compressed to $\SI{44}{\femto\second}$ using an $\text{LG}_{0,3}$ beam, boosting the peak power of an Ytterbium laser system from $\SI{2.5}{\giga\watt}$ to $\SI{9.1}{\giga\watt}$. The spatial homogeneity of the output $\text{LG}_{0,l}$ beams is quantified and the topological charge is spectrally-resolved and shown to be conserved after compression by employing a custom spatio-temporal coupling measurement setup.


\end{abstract}

\maketitle

\section{\label{sec:level1}Introduction}

Ytterbium (Yb)-doped amplified laser systems have proven to be on the cutting edge for efficient high repetition rate - high power laser technology, enabling high-order harmonics generation (HHG) at high flux \cite{klas2020generation} and kHz laser-plasma acceleration (LPA) \cite{farace2023plasma}. However, the narrow spectral bandwidth of these systems leads to pulse durations of several hundreds of femtoseconds (fs), constraining the usable peak powers of such sources. 

In order to overcome this limitation, pulse post-compression methods can be used \cite{nagy2021high}. The multi-pass cell (MPC) technique \cite{Schulte2016, Viotti2022}, based on self-phase modulation (SPM), exhibits high compression ratios, excellent spatio-spectral qualities and high throughput. 
Most MPCs employ a Herriott geometry \cite{Herriott1964}, and consist of two concave mirrors forming a stable resonator that maintains an eigenmode and contains a Kerr medium. The SPM is incremented multiple times along the folded optical path inside the cell, with a non-linear medium that can either be a gas or a bulk material. Gas MPCs are the most recurrent type of cell, allowing the combination of GW-input peak powers and high pulse energies \cite{kaumanns2018multipass, pfaff2023nonlinear} to reach large output peak powers up to the sub-TW regime \cite{Rajhans2023} and few cycle pulse durations \cite{hadrich2022carrier}. With this comes the cost of an extended footprint and the use of rare gases, usually neon or argon, requiring vacuum or high-pressure equipment. 
Bulk MPCs, on the other hand, are less common for large peak power increases due to their apparent lack of energy scalability. They originally operate with MW-input peak powers \cite{grobmeyer2020self} since a bulk material, as opposed to gas, cannot sustain high power or forgive an excess of intensity without getting permanently damaged. A widely-used non-linear medium in bulk MPCs is Fused Silica (FS), which exhibits a critical power for self-focusing of $P_{\text{crit}} =\SI{4.16}{\mega\watt}$ at $\SI{1030}{\nano\meter}$ \cite{Marburger1975}. To reach high peak power outputs in bulk MPCs, the system has to be operated with input powers much greater than $P_{\text{crit}}$ \cite{Raab2022}, which becomes challenging for managing self-focusing and mirror damages. To mitigate these issues, one can reduce the effect of self-focusing by enlarging the beam size, thereby extending the length of the MPC, or by changing the position and thickness of the bulk medium to allow smaller increments of non-linearity \cite{Raab2022}. To comply with the inherent lensing from the Kerr medium, one also needs to use non-linear mode-matching \cite{Hanna2021, Seidel2022} to adapt the beam input parameters to the complete system. 

As a result, the scalability in energy of bulk MPCs is still to be addressed in order to compete with the parameter range of gas MPCs and other post-compression systems, such as gas-filled hollow-core fibers (HCF) \cite{fan202170}. So far, the maximum reported input pulse energy is $\SI{400}{\micro\joule}$ in a hybrid air-bulk Herriott configuration \cite{omar2024hybrid}. To further increase the input energy, multiple approaches can be envisioned, such as convex-concave geometries \cite{Hariton2023, omar2023convex}, hybrid MPCs \cite{Seidel2022, omar2024hybrid}, bow-tie geometries \cite{Heyl2022, schonberg2024compact}, temporal pulse division \cite{stark2022division} and spatial shaping \cite{Kaumanns2021}. The latter technique was demonstrated in a gas MPC by employing a Laguerre-Gaussian $\text{LG}_{0,1}$ mode generated by a spiral phase plate to spread the energy on the transverse plane, showing the compressibility of $\SI{112}{\milli\joule}$, $\SI{1.3}{\pico\second}$ pulses down to $\SI{37}{\femto\second}$ \cite{Kaumanns2021}.

In this work, we extend the approach of spatial shaping to bulk MPCs, using the versatility of a spatial light modulator (SLM) to study high-order Laguerre-Gaussian single-vortex beams ($\text{LG}_{0,l}$, $l=1-3$). We report the post-compression of $\SI{180}{\femto\second}$ pulses with $\SI{610}{\micro\joule}$ pulse energy, corresponding to $\SI{2.5}{\giga\watt}$ input peak power, down to $\SI{44}{\femto\second}$ with high throughput, in a compact bulk MPC with a footprint smaller than $\SI{40}{\centi\meter}$, using a $\text{LG}_{0,3}$ mode. This translates to an output peak power of $\SI{9.1}{\giga\watt}$ with maintained spatial properties. We first describe the non-linear mode-matching conditions and the theoretical energy scaling allowed by $\text{LG}_{0,l}$ beams. Then, we show the experimental realization and characterize the temporal and spatial profiles, showing similar results for all three orders. We also use a custom two-parameter characterization to quantify the spatial homogeneity after the MPC. Finally, we adapt a spatio-temporal couplings (STCs) measurement setup to $\text{LG}_{0,l}$ beams and show that the transverse profile of a $\text{LG}_{0,1}$ mode remains unchanged for different wavelengths. The beam's phase is also spectrally-resolved, demonstrating conservation of the topological charge through post-compression.



\section{\label{sec:level1}Theoretical framework for energy scaling with $\text{LG}_{0,l}$ beams}

All simulations are based on the geometrical characteristics of the experimental setup, further discussed in \ref{labeluseful}.

\subsection{\label{sec:level2}Description of Laguerre-Gaussian beams}

The mathematical form of Laguerre-Gaussian beams is as follows:

\begin{equation}
\begin{split}
    LG_{m,l}\left( \rho, \phi, z\right)  = G \left( \rho, \phi, z \right) \times \left( \frac{\rho}{W(z)} \right) ^{|l|} L_m^{|l|} \left(\frac{2\rho^2}{W^2(z)}\right) \\
     \times \exp \left[- jl\phi + j\left( |l| + 2m\right)\zeta(z) \right]
\end{split}
\label{eq:LGmode}
\end{equation}

\noindent where $G \left( \rho, \phi, z \right)$ represents the amplitude of a Gaussian beam in the cylindrical coordinate system, $L_m^{|l|}$ is the Laguerre polynomial, with $m$, $l$ the radial and azimuthal orders of the considered Laguerre-Gaussian beam, and $\zeta(z)$ is the Gouy phase. As we focus on the use of $m = 0$ radial orders, $\text{LG}_{0,l}$ beams are referred to as vortex of order $l$ in the rest of the study. Both the critical power $P_{\text{crit}}^{(l)}$ for self-focusing \cite{Vuong2006} and the beam size $W^{(l)}(z)$ \cite{Phillips1983} depend on the vortex order, as:

\begin{align}
    P_{\text{crit}}^{(l)} = & \frac{2^{2|l|+1}\Gamma(|l|+1)\Gamma(|l|+2)}{2\Gamma(2|l|+1)}P_{\text{crit}}^{(G)}\label{eq:Pcrit}\\
    W^{(l)}(z) = & \sqrt{|l|+1}W(z) = W_0\sqrt{|l|+1}\sqrt{1+\left(\frac{z}{z_r}\right)^2} \label{eq:beamsize}
\end{align}

\noindent where $\Gamma$ is the Gamma function, $W(z)$ the Gaussian beam size, $W_0$ the waist size and $P_{\text{crit}}^{(G)}$ the critical power for self-focusing of a Gaussian beam. For the same energy contained in the beam, the peak intensity of a vortex beam $I^{(l)}$ will be lower than the Gaussian-equivalent peak intensity $I^{(G)}$, following the law:

\begin{equation}
    \frac{I^{(l)}}{I^{(G)}} = \frac{|l|!}{\left(|l|/e\right)^{|l|}}. \label{eq:intens}
\end{equation}

\begin{figure}[h]
    \centering
    \includegraphics[width=\linewidth]{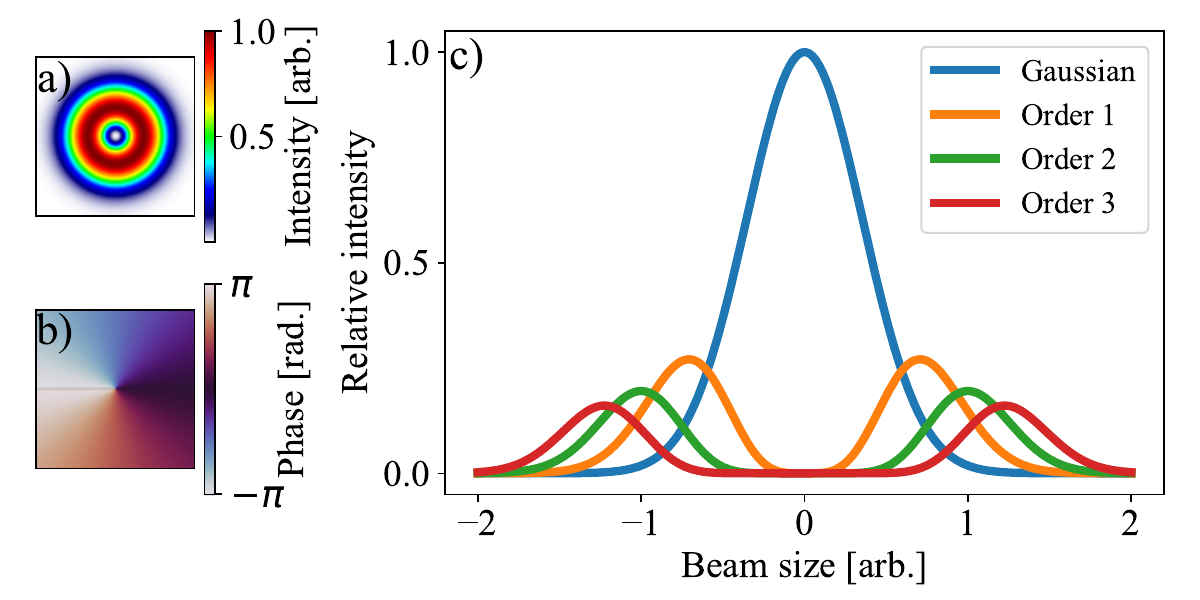}
    \caption{a) Theoretical intensity and b) phase of a vortex of order 1. c) Relative intensities for a Gaussian mode and vortices of order $l=1-3$, with the same pulse energy.}
    \label{fig:th_pic}
\end{figure}

The theoretical intensity and phase maps of a vortex of order 1 are presented in Fig.~\ref{fig:th_pic} a) and b), and a visualization of Eq.~\ref{eq:intens} is provided in Fig.~\ref{fig:th_pic} c). It shows that, for the same pulse energy, the intensity is strongly reduced by using vortex beams compared to a Gaussian profile, and that higher-order vortices incrementally spread over larger spatial dimensions.

\subsection{\label{sec:level2}Non-linear mode-matching with vortices}

To efficiently propagate a laser beam in an MPC, mode-matching is required. In the case of bulk MPCs, lensing due to self-focusing in the bulk medium also needs to be taken into account. This is referred to as non-linear mode-matching (NLMM \cite{Hanna2021, Seidel2022}). Compared to the Gaussian case, self-focusing is not as easily described, due to the orbital angular momentum (OAM) and the equivalent ring lens \cite{Vuong2006}. Yet, this particular self-focusing is still caused by the optical Kerr effect $n = n_0 + n_2I$ (where $n$ is the refractive index of the material, $n_0$ and $n_2$ are the linear and non-linear refractive indices, and $I$ is the intensity of the laser pulse), which is at the core of NLMM. A spatial propagation simulation with ideal vortex beams is performed for the MPC, disregarding temporal effects such as SPM or Group Delay Dispersion (GDD), to find the NLMM of the MPC. Self-focusing is simulated using the non-linear Schrödinger equation, solved using a 4th-order Runge-Kutta (RK4) scheme \cite{runge1895numerical}.
The MPC parameters mimic the experimental configuration, with a total of 14 passes through two $\SI{1}{\milli\meter}$-thick plates of fused silica. A Bayesian Optimization algorithm \cite{BayOpt} is employed to find the beam parameters (focus size and position) that reduce beam size variations on the mirrors from pass to pass. The use of such algorithm is of interest to converge towards the optimal parameters in a small number of steps, even in a multi-dimensional parameter space.
If the cavity is linearly mode-matched, the self-focusing induced by a bulk material will shift the focus negatively, before the linear focus. Thus, the range of optimization given to the algorithm for the focus offset should be in the positive region. The waist size is also expected to vary, but its range is not restricted for the optimization.

\begin{figure}[h]
    \centering
    \includegraphics[width=\linewidth]{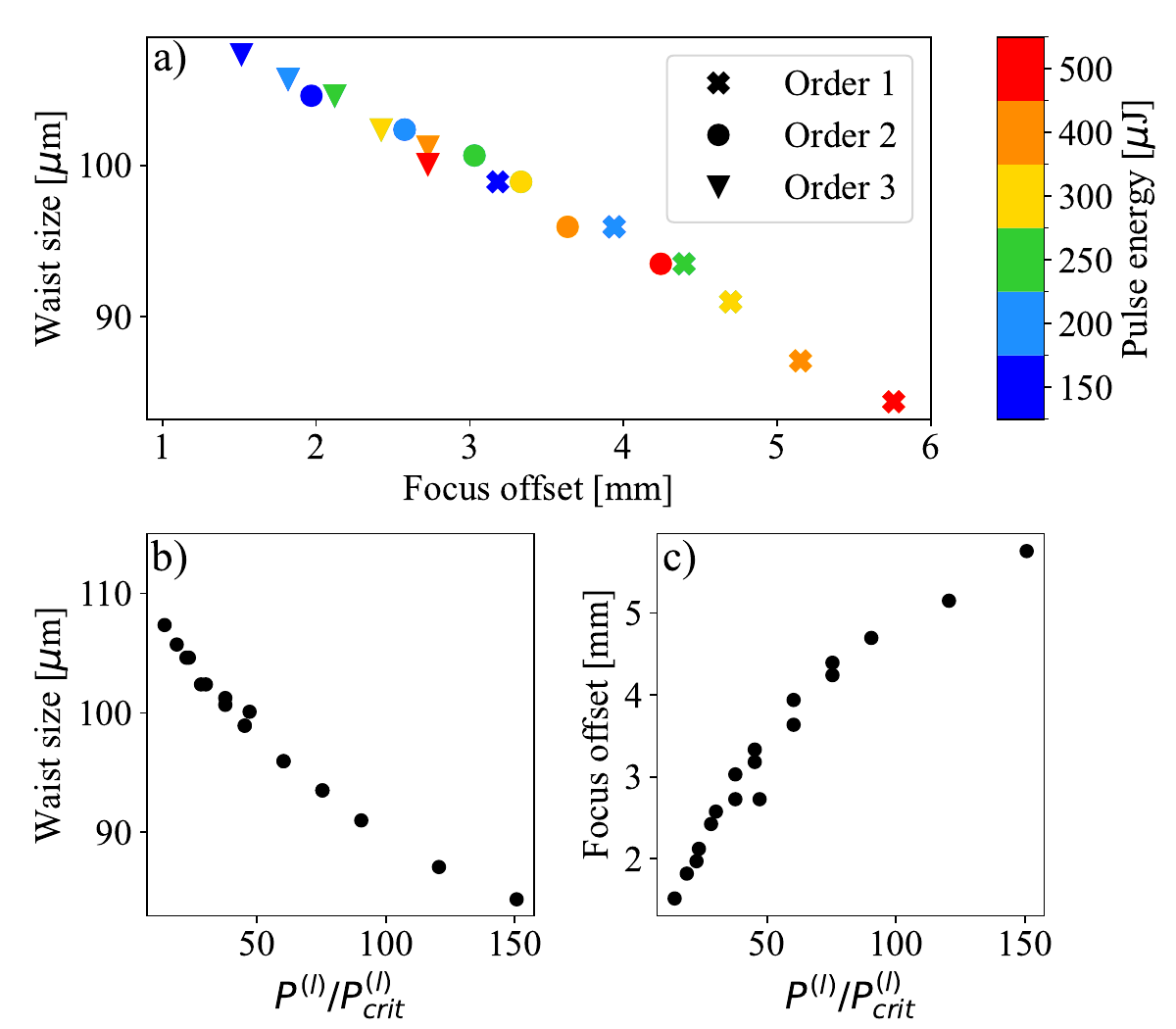}
    \caption{a) Optimized waist size and focus offset of the beam. $\SI{0}{\milli\meter}$ corresponds to the linear mode-matching focus position. The pulse energies are ranging from $\SI{150}{\micro\joule}$ to $\SI{500}{\micro\joule}$. Variation of b) the waist size and c) the offset in focus position in the cell as a function of the ratio of peak power $P^{(l)}$ to critical power $P_{\text{crit}}^{(l)}$ for self-focusing in FS, for the same range of pulse energies.}
    \label{fig:NLMM}
\end{figure}

In Fig.~\ref{fig:NLMM}, the NLMM results are shown for the first three vortex orders, with pulse energies spanning from $\SI{150}{\micro\joule}$ to $\SI{500}{\micro\joule}$. For each order and pulse energy, the waist size and focus offset that minimize the oscillations of the beam size on the mirrors over 14 passes are plotted in Fig.~\ref{fig:NLMM} a). We note that all vortex orders follow the same trend. Figure~\ref{fig:NLMM} b) and Fig.~\ref{fig:NLMM} c) display instead the optimization results as a function of the ratio of peak power to critical power $P^{(l)}/P_{\text{crit}}^{(l)}$ for self-focusing in FS. In this case, clear linear correlations are observed. From this, we conclude that the NLMM in MPCs is not dependent on the vortex order but only on the ratio $P^{(l)}/P_{\text{crit}}^{(l)}$, which gives a recipe for a general mode-matching: by computing $P^{(l)}/P_{\text{crit}}^{(l)}$ for a given vortex order, one can mode-match it to the MPC eigenmode by experimentally realizing the corresponding beam size and focus position obtained from the optimization. This model can then be used to maximize the throughput of the experimental MPC and have a stable mode in the resonator. It is important to notice that these results are valid for one given geometry and dimensions of the system (cell length, position and thickness of the plates), and the same optimization has to be performed for each setup, which would end up in different values yet similar dynamics with respect to $P^{(l)}/P_{\text{crit}}^{(l)}$.

\subsection{\label{scaling_theory}Energy scaling in MPCs with vortices}

The energy scaling factor is defined as the ratio between the vortex pulse energy and the Gaussian pulse energy for which the spectra of the two pulses have the same Fourier Transform Limit (FTL). To derive the energy scaling factor of each order, the RK4 solver is now used in the temporal domain to simulate self-phase modulation, and to extract the broadened spectra. The spatially-averaged normalized spectrum $S_{avg}^{(l)}(E)$ from a spatial intensity profile of an order $l$ vortex is obtained by:

\begin{equation}
        S_{avg}^{(l)}(E,\lambda) = \frac{1}{I^{(l)}_{max}} \iint \mathscr{S}\left(I^{(l)}(x,y,E),\lambda\right)\times I^{(l)}(x,y,E) \,dx \,dy
\end{equation}

\noindent where E is the pulse energy, $\mathscr{S}\left(I^{(l)}(x,y,E),\lambda\right)$ is the normalized broadened spectrum after $14\times\SI{2}{\milli\meter}$ of FS (the total amount seen by the pulse in the experimental setup) for a local intensity $I^{(l)}(x,y,E)$, and $I^{(l)}_{max}$ is the maximum value of the intensity.

\begin{figure}[h]
    \centering
    \includegraphics[width=\linewidth]{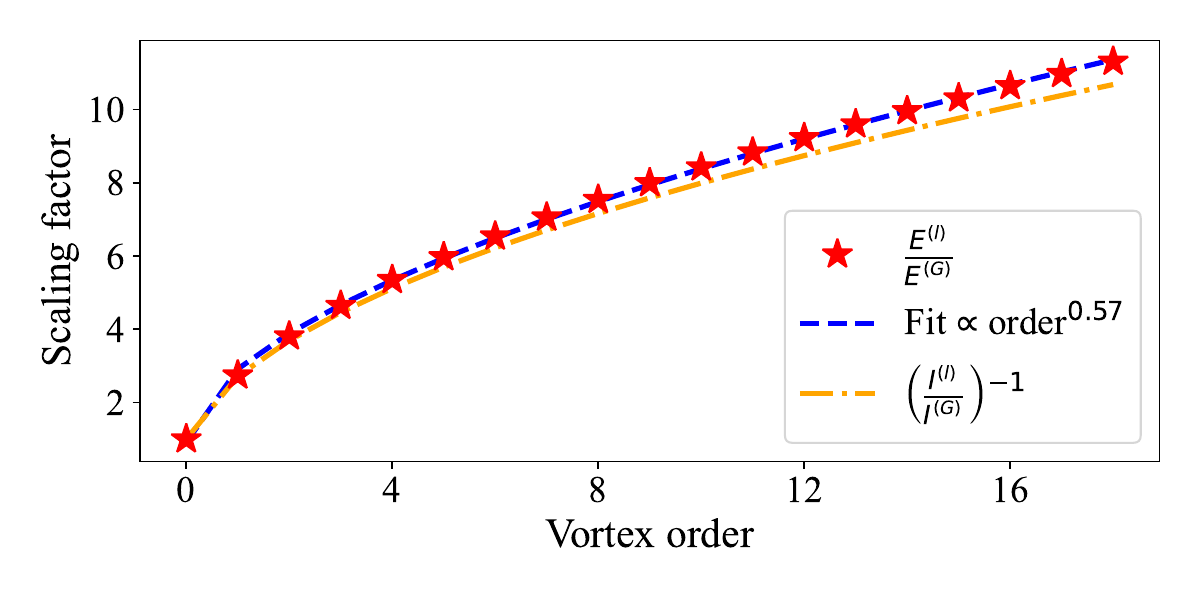}
    \caption{Theoretical scaling derived using Bayesian Optimization to match the Gaussian FTL, from vortex order 0 to 18 (order 0 being a standard Gaussian).}
    \label{fig:thscal}
\end{figure}

The SPM is simulated at different intensities by assuming a plane wave propagator in the MPC, removing all possible spatio-temporal couplings in the laser beam. As the distribution in intensity is dependent on the order of the vortex, we again used Bayesian Optimization to compute the best scaling factor, shown in Fig.~\ref{fig:thscal}. 

The discrete data can be fitted using a power law $l^{0.57}$ (in dashed blue in Fig.~\ref{fig:thscal}). Plotting the inverse of the intensity scaling law (Eq.~\ref{eq:intens}), we see that the fit is close to this other scaling law (which can be approximated as $\sqrt{2\pi l}$). We can intuitively understand this similarity in behavior: SPM is mainly influenced by the peak intensity of the temporal pulse, as it arises from the optical Kerr effect. As the intensity decreases with the vortex order, the effect of SPM will also decrease in the same way, and to compensate for the loss of non-linearity, the energy has to be increased to have the same peak intensity for all orders.
From this power law scaling, one would want to use the highest order possible to increase the energy by a large factor. But as the vortex order increases, so does the size of the beam. Therefore, in practice, pushing towards the highest vortex orders might be limited by the size of the optical components. 

\section{\label{sec:level1}Experimental results}

\subsection{\label{labeluseful}Pulse post-compression setup}

\begin{figure}[h]
    \centering
    \includegraphics[width=\linewidth]{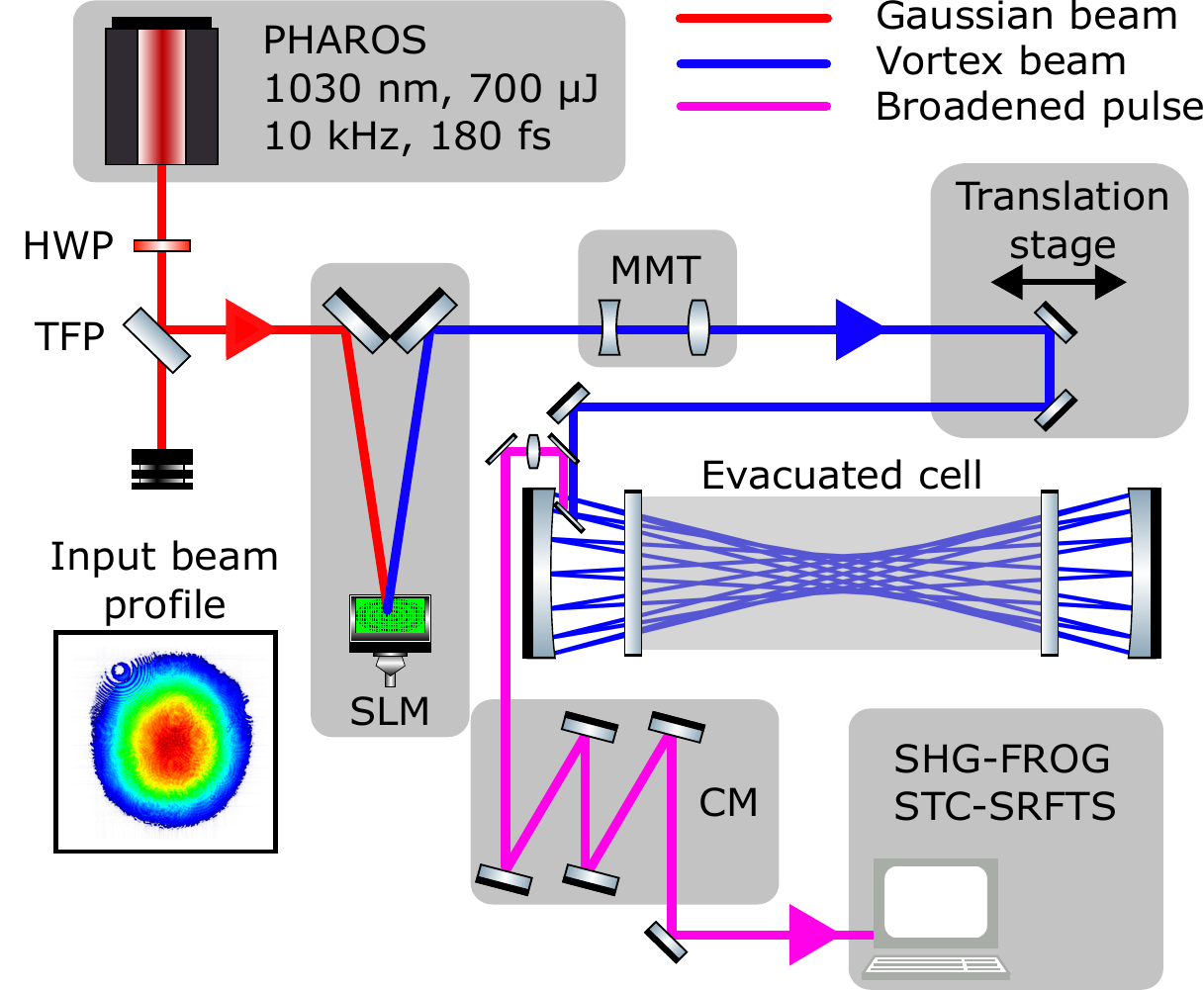}
    \caption{Experimental setup for the use of vortex beams. SLM: Spatial Light Modulator, HWP: Half Wave Plate, TFP: Thin Film Polarizer, MMT: Mode-Matching Telescope, CM: Chirped Mirrors, SHG-FROG: Second Harmonic Generation - Frequency-Resolved Optical Gating, STC-SRFTS: Spatio-Temporal Couplings - Spatially-Resolved Fourier Transform Spectrometry. The initial beam profile is shown in the inset.}
    \label{fig:setup}
\end{figure}

The experimental setup is shown in Fig.~\ref{fig:setup}. The Ytterbium-based laser system is a PHAROS (\textit{Light Conversion}) delivering $\SI{180}{\femto\second}$ Full Width at Half Maximum (FWHM) pulses with energies up to $\SI{700}{\micro\joule}$ and a repetition rate of $\SI{10}{\kilo\hertz}$. The initial beam profile is shown in the inset of Fig.~\ref{fig:setup}. First, the beam is sent to a Spatial Light Modulator (\textit{Hamamatsu Photonics}), used in reflection, to be shaped into a vortex beam. Such device uses liquid crystals to change the local refractive index experienced by the beam, thus imprinting a phase onto it. This way, one is able to generate any vortex beam and correct possible optical aberrations introduced by the setup. 

At the exit of the laser, a manual attenuator consisting of a half-wave plate and a thin film polarizer is placed to study the energy scaling factor. After shaping by the SLM (with less than 2\% losses), the beam is sent through an adjustable mode-matching lens telescope and a translation stage to allow tuning of the propagation distance to the MPC and adjusting the beam parameters following the results of Fig.~\ref{fig:NLMM}. The beam is coupled into the MPC where it undergoes non-linear effects under 7 round-trips, each pass consisting of propagation through two $\SI{1}{\milli\meter}$-thick anti-reflection coated FS plates, separated by $\SI{285}{\milli\meter}$. The MPC mirrors have a $\SI{200}{\milli\meter}$ radius of curvature and are placed $\SI{382}{\milli\meter}$ apart. Between the plates, an evacuated cell (less than $\SI{10}{\milli\bar}$ pressure) is placed to eliminate any non-linear effects from the air. The beam is coupled out of the cell, collimated and compressed using dispersive mirrors. The output compressed pulses are then characterized using in-house diagnostic devices, described in the following sections. 

\subsection{\label{sec:level2}Energy scaling results}

The broadened spectrum of a pulse with a Gaussian spatial intensity profile is recorded after the MPC for three input pulse energies measured after the SLM ($\SI{150}{\micro\joule}$, $\SI{175}{\micro\joule}$ and $\SI{200}{\micro\joule}$). Then, for different vortex orders, the spectral widths are matched to these Gaussian-equivalent references, by tuning the input pulse energy.

\begin{figure}[h]
    \centering
    \includegraphics[width=\linewidth]{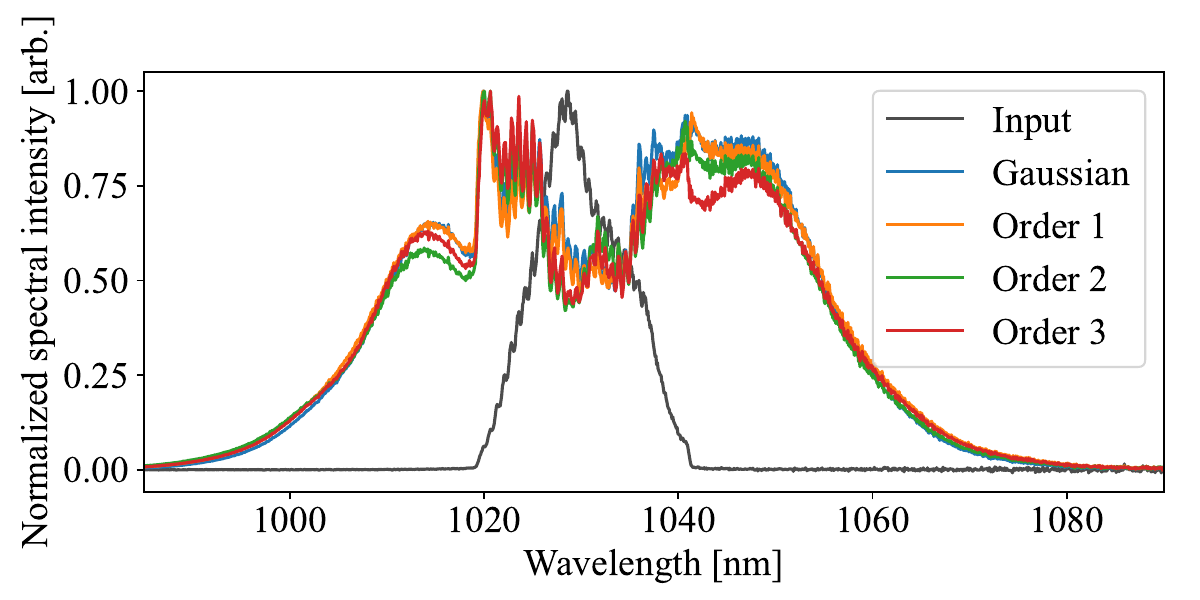}
    \caption{Experimental spectra for a Gaussian-equivalent broadening of $\SI{200}{\micro\joule}$ for $l = 1-3$. The initial spectrum from the laser is shown in black.}
    \label{fig:200spec}
\end{figure}

In Fig.~\ref{fig:200spec}, the spatially-averaged spectra taken for a Gaussian-equivalent broadening at $\SI{200}{\micro\joule}$, and measured using an integrating sphere, are presented. The corresponding input energies are indicated in Table \ref{tab:Escaling}.

\begin{table}[h]
\caption{\label{tab:Escaling}Corresponding input pulse energies and the average energy scaling factor.}
\begin{ruledtabular}
    \begin{tabular}{cccc}
        Gaussian & Order 1 & Order 2 & Order 3 \\ \hline 
        $\SI{150}{\micro\joule}$ & $\SI{270}{\micro\joule}$ & $\SI{343}{\micro\joule}$ & $\SI{450}{\micro\joule}$ \\ 
        $\SI{175}{\micro\joule}$ & $\SI{315}{\micro\joule}$ & $\SI{396}{\micro\joule}$ & $\SI{530}{\micro\joule}$ \\
        $\SI{200}{\micro\joule}$ & $\SI{350}{\micro\joule}$ & $\SI{451}{\micro\joule}$ & $\SI{610}{\micro\joule}$ \\ \hline
        Average energy scaling factor & 1.8 & 2.3 & 3.0 \\ 
    \end{tabular}
\end{ruledtabular}
\end{table}

The scaling shows a significant increase of the input pulse energy that can be sent into the MPC for the same broadening compared to a Gaussian beam, up to a factor 3 using a vortex of order 3. It has to be noted that this scaling is slightly lower than the predicted factor in \ref{scaling_theory}, $\approx 4.5$. 
This is due to the fact that the simulations are performed with an ideal homogeneous vortex beam, free from any localized intensity variations that could induce a stronger spectral broadening. In our experimental setup, because of the SLM segmentation, the beams can carry hotspots, which reduce the energy needed to have the same peak intensity as compared to an ideal vortex (see \ref{spaceCharac}). Consequently, this also affects the non-linear propagation. The downside of using large orders is that the beam size is already doubled with an order 3 vortex. However, the vortex of order 3 still fits on the in- and out-coupling mirror, which is the smallest optical component of the MPC. The transmission of the cell remains above 90\% for all measurements, ensuring that there is no losses due to ionization or clipping.

\begin{figure}[h]
    \centering
    \includegraphics[width=\linewidth]{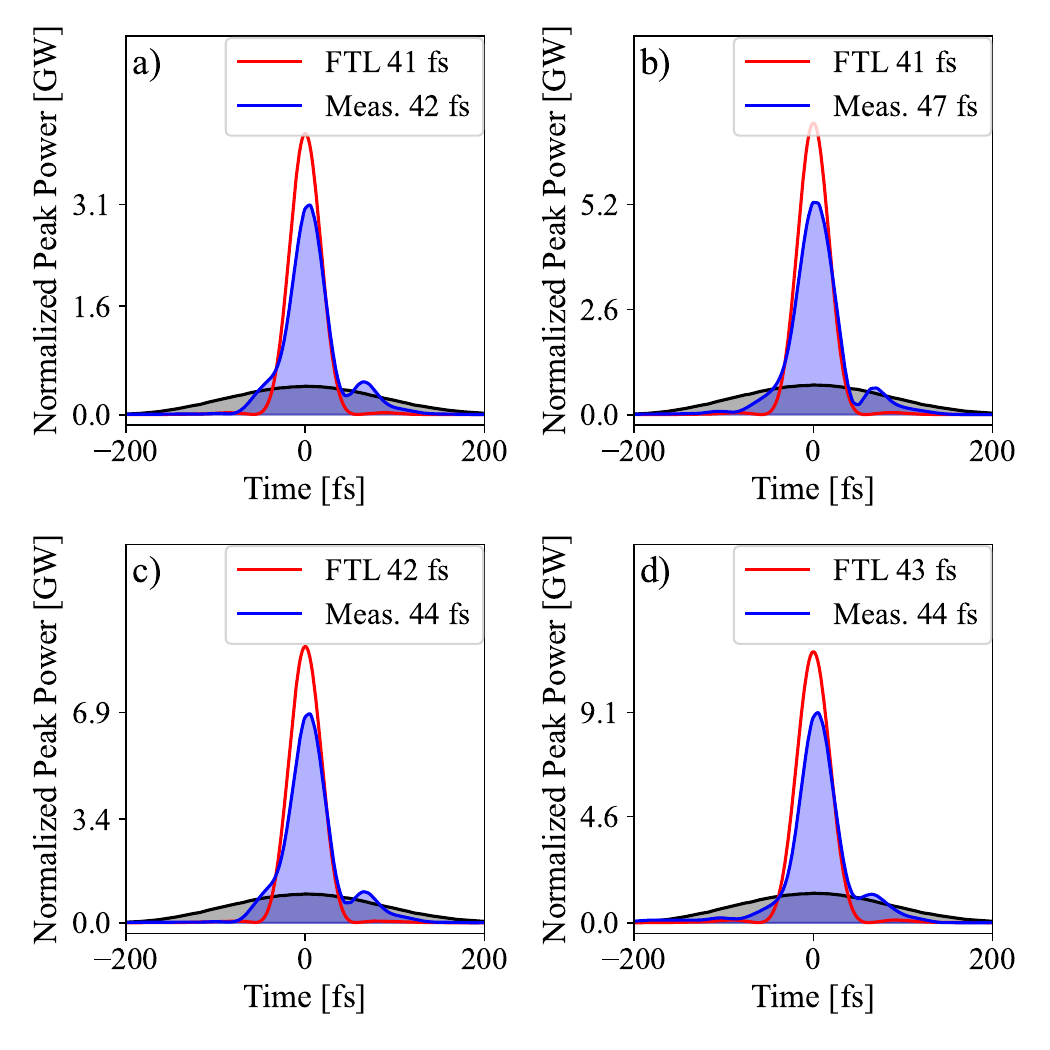}
    \caption{SHG-FROG retrievals (blue) and FTL profiles (red) at $\SI{200}{\micro\joule}$ Gaussian-equivalent broadening for a) Gaussian, b) order 1 vortex, c) order 2 vortex and d) order 3 vortex. The SHG-FROG retrieval of the input laser pulse is shown in black.}
    \label{fig:frogs}
\end{figure}

The MPC output is compressed using 12 bounces on dispersive mirrors, compensating a total GDD of $\SI{2400}{\femto\second}^2$. The resulting pulses are characterized using second harmonic generation frequency-resolved optical gating (SHG-FROG) \cite{kane1993frog} with no spatial sampling. In Fig.~\ref{fig:frogs}, the retrieved temporal profiles corresponding to a $\SI{200}{\micro\joule}$ Gaussian-equivalent spectral broadening are shown. The compression is optimized for the $\SI{200}{\micro\joule}$ Gaussian spatial intensity profile.

From the retrievals we see that all pulses are similar to the Gaussian case, implying that there is no effect on the temporal properties other than the scaling in energy and peak power as observed in Fig.~\ref{fig:frogs}. We retrieve pulses between $\SI{42}{\femto\second}$ and $\SI{47}{\femto\second}$ FWHM, with an FTL around $\SI{42}{\femto\second}$. We also see a clear effect from Third Order Dispersion (TOD), which leads to a side pulse. The TOD is most likely coming from the many reflections on dispersive mirrors, which only allow control of the GDD. Yet, the main pulse contains $>\SI{80}{\percent}$ of the pulse energy, enabling further experimental utilization.

Using the data from Table~\ref{tab:Escaling}, the transmission of the post-compression setup and the FROG retrievals, the peak powers can be estimated. In Table~\ref{tab:PeakP}, the peak powers recorded at $\SI{200}{\micro\joule}$ Gaussian-equivalent broadening are summarized. A maximum output peak power of $\SI{9.1}{\giga\watt}$ was reached for a vortex of order 3, corresponding to a factor 3.6 increase. Providing that the spectral phase could be fully compensated for, one could achieve a peak power of $\SI{11.7}{\giga\watt}$.

\begin{table}[h]
\caption{\label{tab:PeakP}Peak power scaling for equivalent broadenings.}
\begin{ruledtabular}
    \begin{tabular}{ccccc}
         & Gaussian & Order 1 & Order 2 & Order 3 \\ \hline 
         Input peak power (GW) & 0.8 & 1.5 & 1.9 & 2.5 \\ 
        Output peak power (GW) & 3.1 & 5.3 & 6.9 & 9.1 \\ \hline
        Peak power boost & 3.9 & 3.5 & 3.6 & 3.6
    \end{tabular}
\end{ruledtabular} 
\end{table}

\subsection{\label{spaceCharac}Spatial characterization of vortices}

The spatial characterization of vortex beams is not straight-forward, as there is no consensus on how to assess the homogeneity of a ring beam. As the propagation still follows the Gaussian formalism, one could perform an $\text{M}^2$ measurement\cite{fathi2024coherent}, normalized to the theoretical vortex beam size (see Eq.~\ref{eq:beamsize}). Yet, this quantity can be blind to the spatial quality of the considered vortices. We propose hereafter the definition of two parameters, the angular Intensity Homogeneity (IH) and the angular Width Homogeneity (WH), to classify the global Ring Homogeneity (RH) of the generated and propagated beams. 

\begin{figure}[h]
    \centering
    \includegraphics[width=\linewidth]{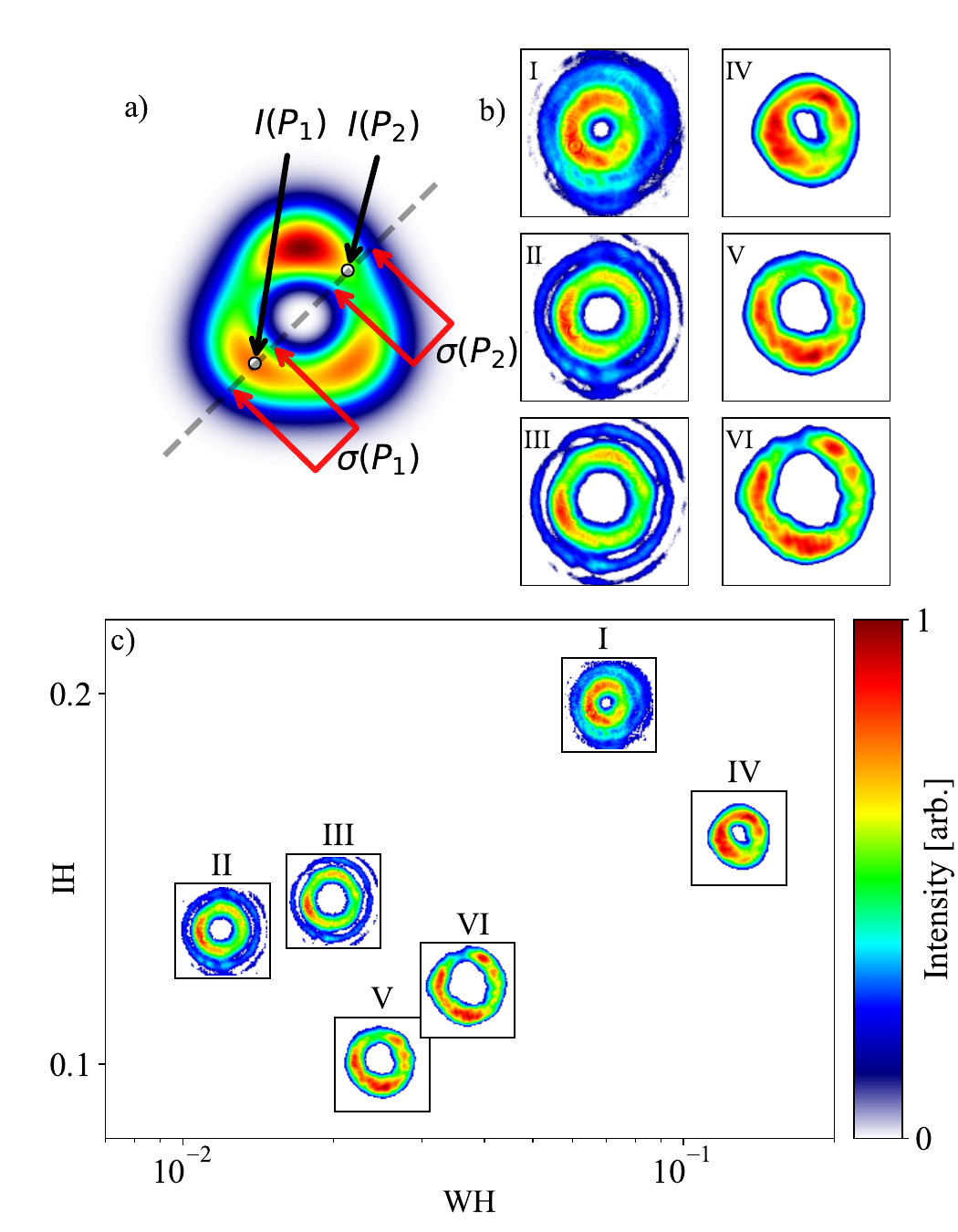}
    \caption{a) Example of a RH characterization on a simulated inhomogeneous vortex, with the relevant parameters only shown over one cut. b) Spatial profiles of the collimated beams before (I: $l = 1$, II: $l = 2$, III: $l = 3$) and after the MPC (IV: $l = 1$, V: $l = 2$, VI: $l = 3$). c) Graphical representation of the RH for the considered profiles. The colorbar representing the normalized intensity is common to all the spatial profiles.}
    \label{fig:exRH}
\end{figure}

First, IH quantifies the variations of intensity maxima on different diameter-cuts through the center of the vortex beam. By taking $n$ cuts over different angles into account, we can define $\text{IH}$ as:

\begin{equation}
    \text{IH} = \frac{1}{n}\sum_{k=1}^n \left|\frac{I(P_{1,k}) - I(P_{2,k})}{I(P_{1,k}) + I(P_{2,k})} \right|, \quad 0 \le \text{IH} \le 1 
\end{equation}

\noindent with $I(P_{1,k})$ and $I(P_{2,k})$ the intensities of the two main peaks over the k-th considered diameter of the ring. This operator is blind to the spread of each peak, that can come from aberrations. Thus, we define WH as:

\begin{equation}
    \text{WH} = \frac{1}{n}\sum_{k=1}^n \left|\frac{\sigma(P_{1,k}) - \sigma(P_{2,k})}{\sigma(P_{1,k}) + \sigma(P_{2,k})} \right|, \quad 0 \le \text{WH} \le 1 
\end{equation}

\noindent where $\sigma(P_{1,k})$ and $\sigma(P_{2,k})$ are the FWHM of the peaks over the k-th considered diameter. In Fig.~\ref{fig:exRH}~a), an example of the characterization is shown, highlighting the considered parameters.

Using these two parameters, the homogeneity of $\text{LG}_{0,l}$ beams is quantified as RH = (IH;WH), with the ideal beam tending towards (0;0). In Fig.~\ref{fig:exRH}~b), the spatial profiles of the beams are shown before and after a $\SI{200}{\micro\joule}$ Gaussian-equivalent broadening in the MPC, for the three first vortex orders. The value of RH are indicated in Table~\ref{tab:RH_table} for each beam profile, and a graphical representation is shown in Fig.~\ref{fig:exRH}~c).

\begin{table}[h]
\caption{\label{tab:RH_table}Ring Homogeneity parameters for input and output beams.}
\setlength\extrarowheight{3pt}
\begin{ruledtabular}
    \begin{tabular}{cccc}
        Order & Position & IH & WH \\ 
        \hline 
        1 & In & $\num{2.0d-1}$  & $\num{7.1d-2}$ \\ 
        & Out & $\num{1.6d-1}$ & $\num{1.3d-1}$ \\ 
        \hline
        2 & In & $\num{1.4d-1}$ & $\num{1.2d-2}$ \\ 
        & Out & $\num{1.0d-1}$ & $\num{2.5d-2}$ \\ 
        \hline
        3 & In & $\num{1.4d-1}$ & $\num{2.0d-2}$ \\ 
        & Out& $\num{1.2d-1}$ & $\num{3.7d-2}$  \\ 
    \end{tabular}
\end{ruledtabular}
\end{table}

After shaping by the SLM, the vortices, shown in Fig.~\ref{fig:exRH} b) I, II and III, exhibit a large diffraction pattern (in blue) due to the phase singularity on the SLM. This pattern disappears after propagation through the MPC, as displayed in Fig.~\ref{fig:exRH} b) IV, V and VI. The IH decreases while the WH slightly increases after spectral broadening. Yet, the vortices are still consistent with their non-broadened spatial profiles, which ensures that the propagation through the MPC is maintaining satisfactory spatial properties for further applications.

\subsection{\label{sec:level2}Spatio-temporal characterization of vortices}

\begin{figure}[h]
\centering
\includegraphics[width=\linewidth]{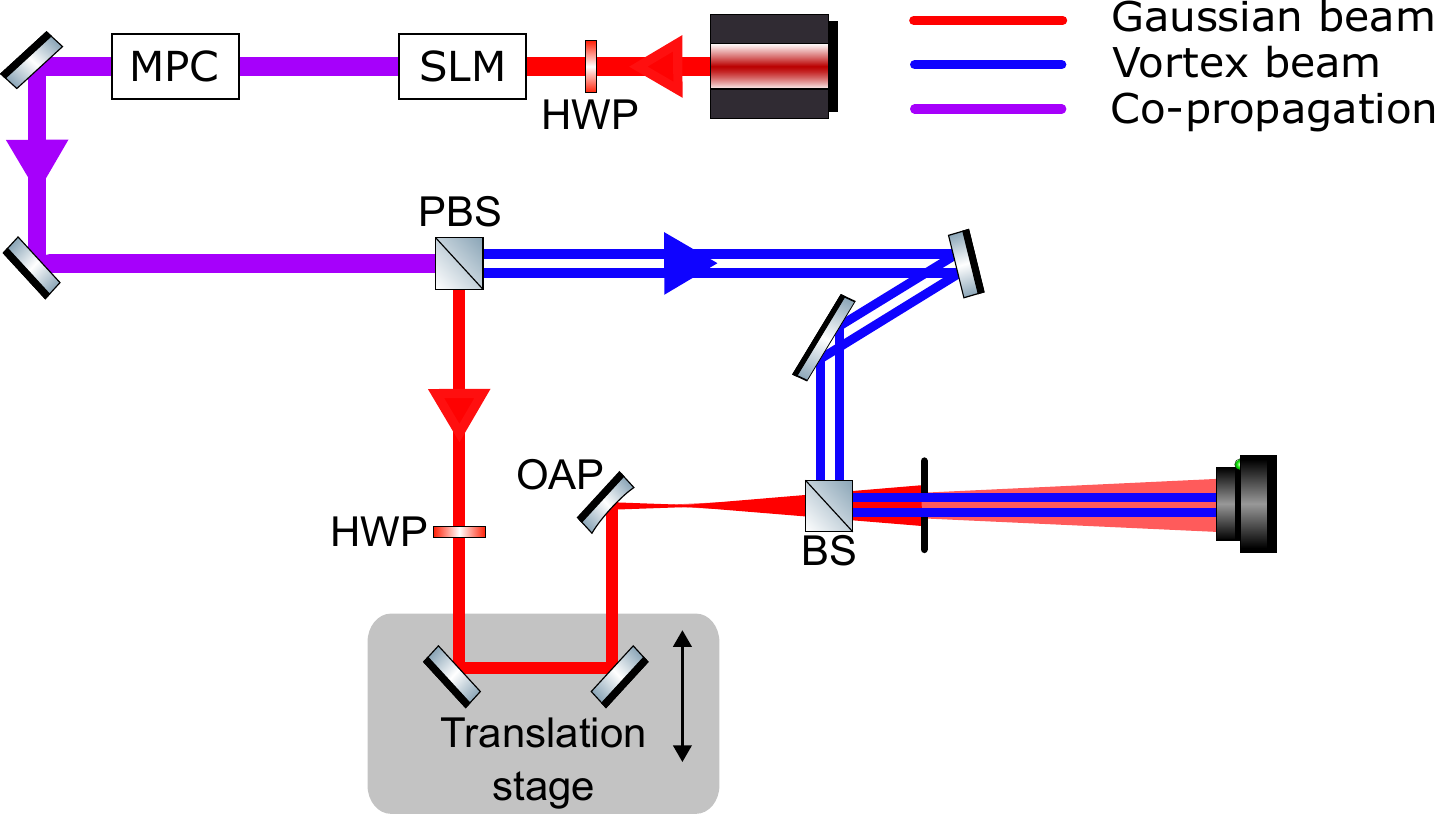}
\caption{Experimental spatio-temporal measurement setup. PBS: Polarization Beam Splitter, BS: Beam Splitter, HWP: Half-Wave Plate, OAP: Off-Axis Parabola.}
\label{fig:SetupSTC}
\end{figure}

A spatio-temporal characterization of the pulses is performed using spatially-resolved Fourier transform spectrometry \cite{miranda2014spatiotemporal}. The setup normally uses a reference beam that is generated by focusing the same laser into a tight spot. Its central part is then sampled as reference for the measurement, which is not possible with a vortex that carries a hole. Instead, we use a particular property of SLMs, which is that they only work for a specific polarization state (in our case S-polarization). The input polarization can be rotated so that the S-polarized component experiences the phase imprint from the SLM to form a vortex beam, while the P-polarized component does not experience a phase imprint and therefore remains Gaussian. By tuning the input polarization angle, the energy ratio between the two beams is changed so that both pulses with different spatial profiles undergo the same amount of spectral broadening in the MPC. The experimental setup is shown in Fig.~\ref{fig:SetupSTC}. After the post-compression setup, the pulse with a vortex beam and the reference pulse with a Gaussian beam are split by a polarization beam splitter. The Gaussian profile gets its polarization rotated to match that of the vortex beam, is then sent to a delay stage and focused with an off-axis parabola to generate a spherical wavefront (a small portion of which is later selected via a pinhole). The two pulses are then recombined by a second beam splitter, and their interference pattern is recorded on a camera for different delays.

As a proof of concept, we measure the spatio-temporal profile of a vortex of order 1, at a Gaussian-equivalent broadening of $\SI{150}{\micro\joule}$, i.e. $\SI{150}{\micro\joule}$ for the Gaussian beam and $\SI{270}{\micro\joule}$ for the vortex beam, for a total of $\SI{420}{\micro\joule}$ input pulse energy. A higher pulse energy could be used to maximize the spectral broadening, but was not considered at this time, as the goal was to simply test the method with relaxed pulse parameters. For further experimental studies utilizing the post-compressed pulses, spatio-temporal characterization will be performed at full energy.

\begin{figure}[h]
\centering
\includegraphics[width=\linewidth]{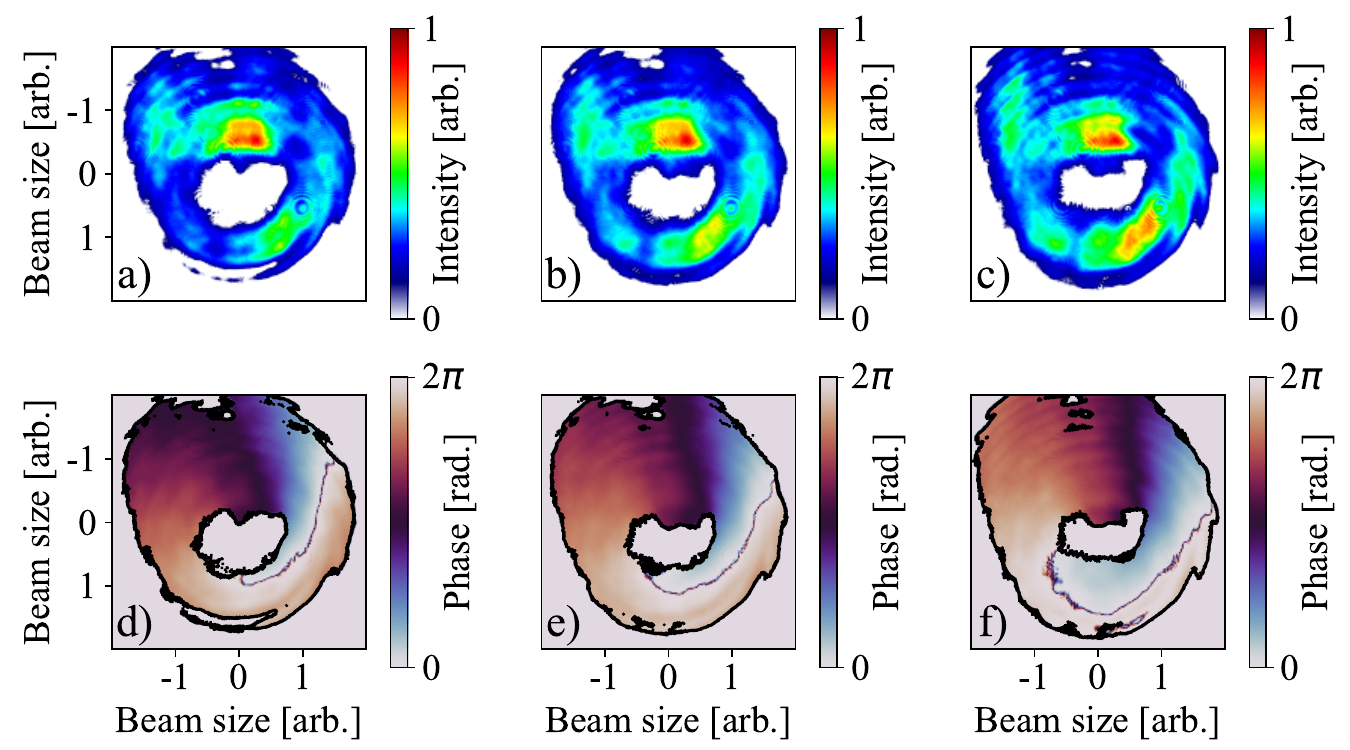}
\caption{Intensity and phase distributions for $\lambda$ = $\SI{1010}{\nano\meter}$ [a) and d)], $\SI{1030}{\nano\meter}$ [b) and e)], $\SI{1050}{\nano\meter}$ [c) and f)].}
\label{fig:STCdata}
\end{figure}


The underlying assumption for this measurement to be valid is that the only non-linear effect acting on the pulses is the SPM. However, due to the co-propagation of two waves with similar intensities, a discussion is needed regarding potential cross-phase modulation (XPM). First, the two pulses do not share the same polarization and they are not temporally overlapped, as the SLM is pseudo-birefringent (only one polarization sees the phase pattern). Secondly, the maxima of the beams are not spatially overlapping, as a vortex beam is by definition hollow. The comparison of the broadened spectra after the MPC for a vortex-only beam and co-propagating vortex and Gaussian beams leads to very little disparities, which indicates that spectral broadening is still dominated by the SPM. For higher energies, the addition of a proper birefringent crystal would ensure a temporal separation large enough to avoid any XPM.

From the STC measurement, the intensity and phase distributions at three different wavelengths are extracted and displayed in Fig.~\ref{fig:STCdata} a)-f), for $\lambda$ = $\SI{1010}{\nano\meter}$ (a) and d)), $\SI{1030}{\nano\meter}$ (b) and e)) and $\SI{1050}{\nano\meter}$ (c) and f)). The three intensity profiles are close to identical, and the three phases exhibit a spiral pattern that is highly similar for all wavelengths. The average ring homogeneity is (0.11;0.78), which is far from the values at the output of the MPC in Table \ref{tab:RH_table}. The WH is especially high, due to a measurement artifact that is seen at the top of the profiles. In our case, shaping the beam of a narrowband pulse with an SLM before spectral broadening allows mitigating losses arising from spectrally-dependent phase imprint. Thus, short pulses can be generated with an "achromatic OAM" for all wavelengths.

\section{\label{sec:level1}Conclusions}

Using a Laguerre-Gaussian single-vortex intensity profile inside a $\SI{382}{\milli\meter}$ long bulk multi-pass cell enables the pulse energy scaling of such post-compression method with a factor 1.8 for vortex of order 1 and up to a factor 3 for an order 3, compared to propagation with a Gaussian spatial profile. Pulses with $\SI{180}{\femto\second}$ duration and $\SI{610}{\micro\joule}$ pulse energy, corresponding to $\SI{2.5}{\giga\watt}$ input peak power, are compressed down to $\SI{44}{\femto\second}$, with a spatial homogeneity suitable for further experimental use. Despite the somewhat large peak powers employed in the bulk MPC, the spatial distribution is unchanged across the broadened spectrum. Moreover, the topological charge is shown to be conserved after non-linear propagation and kept constant across the spectrum. The utilization of an SLM, aside from shaping the beam and making the transition between vortices of different orders extremely simple, allows the precise control of the input intensity and phase profile, something that is not possible with simple spiral phase plates. Aberrations are also corrected to optimize the output beam quality and maximize the throughput of the MPC. 
The output peak power reached in such compact setup is $\SI{9.1}{\giga\watt}$, which is, to the best of our knowledge, a state-of-the-art result for a single-stage Herriott-type bulk MPCs, similar to recent results \cite{Liu2024}, with however $>\SI{500}{\micro\joule}$ pulse energies. While this is still far from the record peak powers attained in gas-filled MPCs \cite{Rajhans2023} and HCF \cite{fan202170}, the compactness and low complexity of the setup is attractive for laser systems with more relaxed parameters. In our experiment, the maximum pulse energy available from the source ($\SI{700}{\micro\joule}$) could almost be entirely used, and we foresee that the scaling technique presented here should easily enable compression of mJ-level pulses with higher-order vortices. Though, the increased beam size of vortices implies the use of larger optics, or possibly segmented optical components, which goes against the idea of a simple, compact experimental setup. In this case, spatial beam shaping can be combined with other innovative cell geometries to reduce the overall footprint \cite{Heyl2022, schonberg2024compact}. Efficiently removing the vortex geometry with the increased bandwidth is also a challenge for experiments requiring a Gaussian-like profile. Phase plates and SLMs can undo the OAM, but the reconstruction of a Gaussian profile would call for additional design. Nevertheless, the direct use of vortices is highly interesting for, e.g., HHG \cite{Carlos2017} and LPA \cite{song2023characteristics}. In particular, high resolution ptychography benefits from OAM beams carrying more phase information in retrievals \cite{wang2023ptycho, hernandez2017review}.

\section*{Fundings}

The authors acknowledge financial support from the Swedish Research Council (Grants Nos. 2021-04691 and 2022-03519) and the Crafoord Foundation.


\begin{acknowledgments}
The authors would like to thank Peter Smorenburg and David O’Dwyer from ASML for fruitful discussions and helpful suggestions during the writing of the paper.
\end{acknowledgments}

\section*{Disclosures}

The authors declare no conflicts of interest.

\section*{Author contributions}

\textbf{V. K.}: Conceptualization (main); Data curation (main); Formal analysis (main); Investigation (main); Methodology (equal); Software (main); Validation (equal); Visualization (main); Writing – original draft (main). 
\textbf{S. W.}: Data curation (equal); Formal analysis (equal); Investigation (equal); Methodology (equal); Validation (equal); Writing – review and editing (supporting).
\textbf{M. R.}: Methodology (equal); Investigation (supporting); Software (supporting); Writing – review and editing (supporting).
\textbf{G. B.}: Investigation (supporting); Writing – review and editing (supporting).
\textbf{A.-K. R.}: Methodology (supporting); Writing – review and editing (supporting).
\textbf{C. G.}: Methodology (supporting); Writing – review and editing (supporting).
\textbf{C. L. A.}: Methodology (supporting); Writing – review and editing (supporting).
\textbf{A.-L. V.}: Conceptualization (main); Funding acquisition (main); Investigation (equal); Methodology (equal); Project administration (main); Supervision (main); Writing – review and editing (equal).

\section*{Data Availability Statement}

The data that support the findings of this study are available from the corresponding author upon reasonable request.

\section*{References}

\bibliography{paper}

\end{document}